\documentclass[runningheads]{llncs}
\usepackage{graphicx}
\usepackage[utf8]{inputenc} 
\usepackage[T1]{fontenc}    
\usepackage{hyperref}       
\usepackage{url}            
\usepackage{booktabs}       
\usepackage{amsfonts}       
\usepackage{nicefrac}       
\usepackage{microtype}      
\usepackage{xcolor}         

\usepackage{amsmath}
\usepackage{graphicx} 
\usepackage{siunitx}

\usepackage[numbers, sort, compress]{natbib}
\usepackage{enumitem}
\usepackage{multirow}
\usepackage{pifont}
\graphicspath{ {figures/} }

\makeatletter
\newcommand{\printfnsymbol}[1]{%
  \textsuperscript{\@fnsymbol{#1}}%
}

\begin{document}
\title{Physics Informed RNN-DCT Networks for Time-Dependent Partial Differential Equations}
\titlerunning{Physics Informed RNN-DCT Networks}
%
\author{Benjamin Wu\thanks{Equal contribution.}\:\thanks{Work done during NVIDIA AI Research Residency}\inst{1,2}
\and
Oliver Hennigh\inst{1}
\and
Jan Kautz\inst{1}
\and 
Sanjay Choudhry\inst{1}
\and 
Wonmin Byeon\printfnsymbol{1}\inst{1}
}
\authorrunning{B. Wu et al.}
%
\institute{NVIDIA, Santa Clara CA 95051, USA\\
\email{\{ohennigh,jkautz,schoudhry,wbyeon\}@nvidia.com} \and
National Astronomical Observatory of Japan, Mitaka Tokyo 181-8588
\email{benwu.astro@gmail.com}}
\maketitle              
\begin{abstract}
Physics-informed neural networks allow models to be trained by physical laws described by general nonlinear partial differential equations. However, traditional architectures struggle to solve more challenging time-dependent problems due to their architectural nature. In this work, we present a novel physics-informed framework for solving time-dependent partial differential equations. Using only the governing differential equations and problem initial and boundary conditions, we generate a latent representation of the problem’s spatio-temporal dynamics. Our model utilizes discrete cosine transforms to encode spatial frequencies and re-current neural networks to process the time evolution. This efficiently and flexibly produces a compressed representation which is used for additional conditioning of physics-informed models. We show experimental results on the Taylor-Green vortex solution to the Navier-Stokes equations. Our proposed model achieves state-of-the-art performance on the Taylor-Green vortex relative to other physics-informed baseline models.
\end{abstract}
\section{Introduction}
\label{sec:intro}

Numerical simulations have become an indispensable tool for modeling physical systems, which in turn drive advancements in engineering and scientific discovery. However, as the physical complexity or spatio-temporal resolution of a simulation increases, the computational resources and run times required to solve the governing partial differential equations (PDEs) often grow drastically.


\textbf{ML-driven solvers. }
Recently, machine learning approaches have been applied to the domain of physical simulation to ameliorate these issues by approximating traditional solvers with faster, less resource-intensive ones. These methods generally fall into two main paradigms: data-driven supervision \cite{Guo2016ConvolutionalNN,bhatnagar2019prediction,hennigh2017lat,zhu2018bayesian,li2020fourier} or physics-informed neural networks (PINNs) \cite{yu2017deep,raissi2019physics,bar2019unsupervised,smith2020eikonet}. \emph{Data-driven approaches} excel in cases for which reliable training datasets are available and if the underlying physical equations are unknown. These generally utilize deep convolutional neural networks to parameterize the solution operator.
\emph{PINN-based solvers} parameterize the solution function directly as a neural network. 
This is typically done by passing a set of query points through a feed-forward fully-connected neural network (or multilayer perceptron, MLP) and minimizing a loss function based on the governing PDEs, initial conditions (ICs) and boundary conditions (BCs). This allows the simulation to be constrained by physics alone and does not require any training data. 
However, the accuracy of traditional PINN-based approaches is limited to problems in low dimensions and with simpler time-independent physics. 

\textbf{Learning time-dependent problems with PINNs.} 
For time-dependent problems, traditional PINN-based models use a continuous time approach which treats the time dimension in the same manner as spatial dimensions. 
To avoid increasingly poor performance as the simulation evolves in time, methods have been developed to split the domain into many short-time problems and solve each step using continuous-time PINNs \cite{meng2020ppinn,mattey2021physics}. 
However, this results in additional model complexity and computational overhead. 
Another approach to learn spatio-temporal domains is the generation of a grid of latent context vectors to condition the spatio-temporal query points entering the MLP. \citet{esmaeilzadeh2020meshfreeflownet} generate latent context grids from low-resolution data using convolutional encoders to perform super-resolution on fluid simulations. 
In principle, a well-constructed latent context grid allows the PINN to learn more easily while still relying on physics-constrained losses. In this work, we design a novel physics-informed MLP architecture by adding a new latent context generation process to effectively learn spatial-temporal physics problems. 

\textbf{Efficient learning in time and space. }
Typical feedforward neural networks lack notions of temporal relationships. Recurrent neural networks (RNNs) form graphs directed along a temporal sequence, which allow for learning of time-dependent dynamics. 
Long Short-Term Memory (LSTM) \cite{hochreiter1997long} and Gated Recurrent Units (GRUs) \cite{cho2014learning} provide a gating mechanism to solve the problem of vanishing gradients and have become popular choices for RNNs, exhibiting high performance and efficiency.
Regarding spatial features, \citet{xu2020learning} demonstrate that compressing high-resolution images in the frequency domain using digital signal processing methods improved model accuracy while greatly reducing input size. 
This allows for high model efficiency while maintaining compatibility with standard spatial CNN architectures.

\textbf{Contributions.} Although PINN solvers provide a well-principled, machine learning approach that promises to revolutionize numerical simulations, their current constraints to problems with simple geometries and short times severely limits their real-world impact. 
We address these shortcomings by introducing novel design choices that improve the simulation accuracy and efficiency of PINN solvers on more challenging problems, particularly in the regime of long time evolution where current PINNs severely struggle. 

Our key contributions are as follows:
\begin{itemize}[label=\ding{212}]
\item We propose a new approach for latent context generation that requires no additional data and enables PINNs to learn complex time-dependent physics problems. 
\item Our work is the first to directly address space-time-dependent physics using PINNs. This is achieved by utilizing convolutional GRUs for learning the spatio-temporal dynamics of simulations. 
\item We separate the spatial and frequency domains, adding flexibility for the network to learn more diverse physical problems.
\item We test the new model against other architectures on benchmark transient simulation problems and demonstrate quantitative improvements in both accuracy and speed. 
\end{itemize}

\section{Related Works}
\label{sec:related}
Traditional PINNs \cite[e.g.,][]{raissi2019physics} find solutions to PDEs of the form
\begin{equation}
    \partial_{t}u + \mathcal{N}[u] = 0, x\in \Omega, t\in [0,T],
\end{equation}
where $u(t,x)$ is the PDE solution function, $\mathcal{N}[\cdot]$ is the (nonlinear) differential operator, and the spatial domain $\Omega \subset \mathbb{R}^{D}$. The continuous time approach defines a PINN $f(t,x)$ such that
\begin{equation}
    f := \partial_{t}u + \mathcal{N}[u].
\end{equation}
As a result, $u(t,x)$ and $f(t,x)$ can be approximated by deep neural networks with parameters optimized by minimizing the mean squared error (MSE) loss:
\begin{equation}
\begin{aligned}
    MSE &= MSE_{u} + MSE_{f} \\
    &= \frac{1}{N_{u}}\sum_{i=1}^{N_{u}}{|u(t^{i}_{u},x^{i}_{u})-u^{i}|^{2}} + \frac{1}{N_{f}}\sum_{i=1}^{N_{f}}{|f(t^{i}_{f},x^{i}_{f})|^{2}},
\end{aligned}    
\end{equation}
where $MSE_{u}$ depends on the points $\{ t_{u}^{i}, x_{u}^{i}, u_{i}\}^{N_{u}}_{i=1}$ based on the state, boundary condition (BC), and initial condition (IC), while $MSE_{f}$ is calculated from $\{t_{f}^{i}, x_{f}^{i}\}^{N_{u}}_{i=1}$ which represent points from $f(t,x)$ in the spatio-temporal domain.

The continuous time approach is effective for short-term integration, but requires a large $N_{f}$ to properly enforce the physical constraints. For long-time integration of PDEs and problems in which additional data is limited or nonexistent, these become significant limitations. As a workaround, \citet{raissi2019physics} introduced discrete time models that borrowed from classical Runge-Kutta methods. By splitting the problem into $q$ discrete time steps, continuous time approaches can be used for each short-time integration step and then combined via classical numerical approaches. This hybrid model enables much larger time steps than those required by traditional Runge-Kutta stepping schemes, but loses the flexibility of purely neural approaches and does not solve the underlying difficulty of the network in learning longer time-dependent problems.

Subsequent work has also taken the approach of decomposing a long-time problem into many short-time problems. \textsc{PPINN}\cite{meng2020ppinn} was a method introduced that uses a fast coarse-grained solver to approximate a simplified PDE and then use those solutions as IC and BC for the independent sub-problems. These continuous time models are supervised in parallel, boosting efficiency and accuracy. 

\textsc{bc-PINN}\cite{mattey2021physics} also uses time segmentation, but requires only one neural network. Instead of using a traditional numerical method to recombine independent problems, the neural network is retrained over each successive time segment while satisfying the solution for all previous time segments. A continuous solution is found despite the time segmentation and accuracy for higher order time-dependent problems was improved.

Our proposed method treats the time domain in a fundamentally different manner, modeling temporal relationships using RNNs. 
Efficiency and accuracy are further improved by learning features in the spatial frequency domain.
These spatial-temporal representations are combined to generate latent context grids, which condition the PINN without the need for additional data.
Overall, our fully neural approach provides a continuous solution while avoiding the limitations and overhead of combining explicit time splitting schemes with continuous time models. 

\section{Methods}
\label{sec:methods}

\begin{figure}
  \centering
  \includegraphics[width=1.0\linewidth]{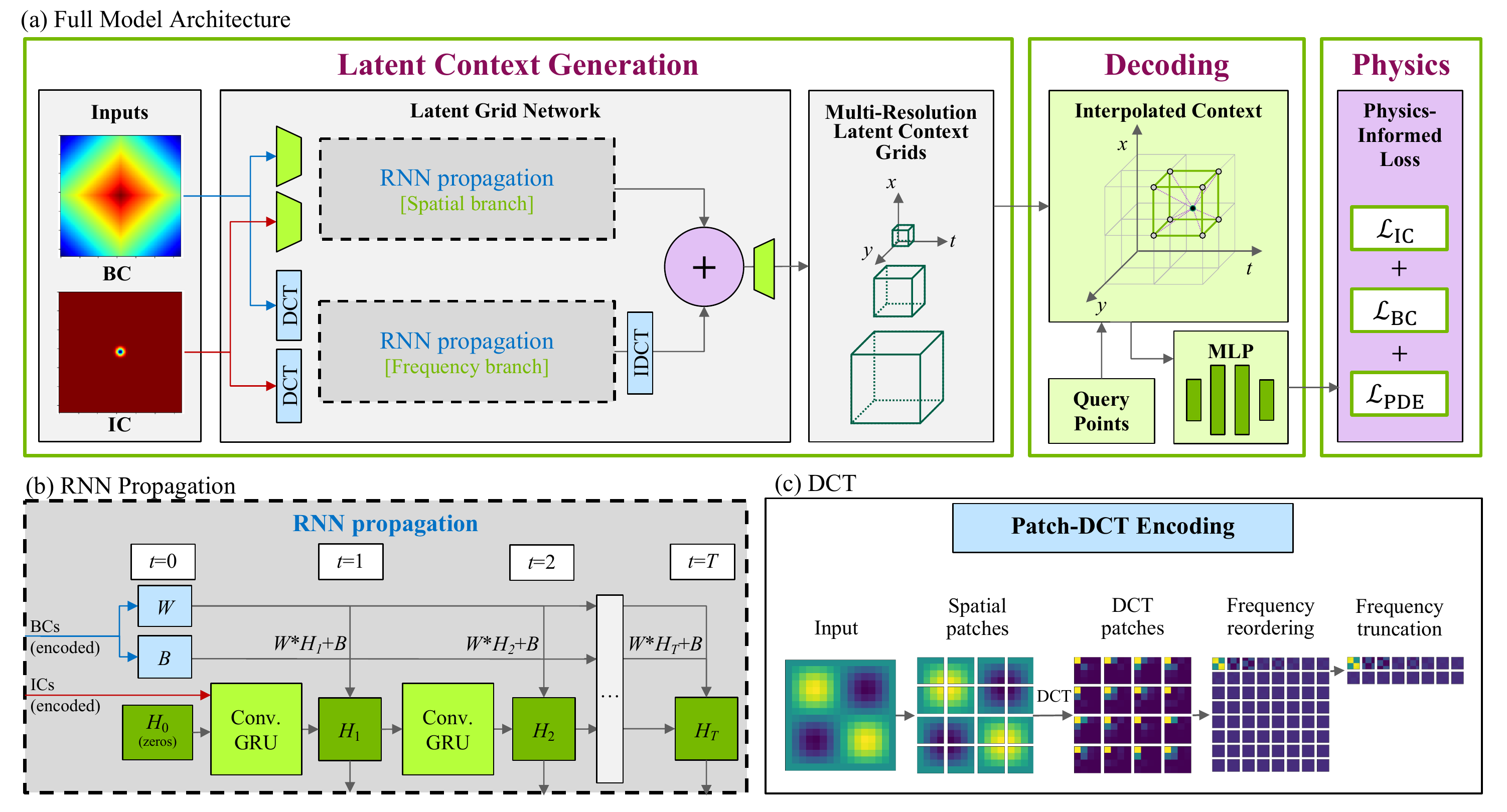}
  \caption{
    (a) Full model architecture. In the latent context generation phase, the problem inputs (IC and BC) are each passed through a latent grid network consisting of two branches of RNN propagation. The structure of the RNN propagation is shown in (b). The inputs pass through convolutional encoders in the spatial branch and patch-DCT encoders (c) in the frequency branch. The outputs are combined into multi-resolution spatio-temporal latent context grids. These grids are used to interpolate the problem query points and provide additional context during a continuous decoding phase through the MLP. Finally, the network is trained in a physics-informed manner using the IC, BC, and PDE losses.
    }
  \label{fig:architecture}
\end{figure}

In this paper, we propose a new model that enables PINN-based neural solvers to learn temporal dynamics in both the spatial and frequency domains. Using no additional data, our architecture can generate a latent context grid that efficiently represents more challenging spatio-temporal physical problems. 

Our full architecture is shown in ~\autoref{fig:architecture} (a). 
It consists of three primary parts, which are explained in more detail below: (1) latent context generation, (2) decoding, and (3) physics-informing. The latent context generation stage takes as input the problem ICs and BCs and outputs spatio-temporal latent context grids. For the decoding stage, spatio-temporal query points along with additional vectors interpolated from the latent context grid are used as input. For each set of points, the MLP predicts corresponding output values, on which the physics-constrained losses are applied. Upon minimization of these losses, the MLP approximates the function governed by the underlying PDEs.



\subsection{Latent grid network}

The primary novelty of our method is the latent grid network that can generate context grids which efficiently represent the entire spatio-temporal domain of a physical problem without the need for additional data. 

This network requires two inputs for the problem-specific constraints: ICs and BCs. The ICs are defined as $u_{0}=u(x_{1,..,N},t=0)$ for each PDE solution function $u$ over $N$ spatial dimensions. The BCs are defined based on the geometry of the problem for each spatial dimension. An additional spatial weighting by signed distance functions (SDFs) can also be applied to avoid discontinuities at, e.g., physical boundaries, but would not be necessary for, e.g., periodic BCs.
Each tensor undergoes an encoding step in either the frequency or spatial domain. 

\textbf{Frequency branch. } 
The {frequency branch} transforms the spatial inputs to frequencies via the discrete cosine transform (DCT), motivated by \cite{xu2020learning}. 
\autoref{fig:architecture} (c) illustrates our patch-wise DCT encoding step. 
First, the ICs and BCs are separately split into spatial patches of size $p\times p$. DCTs are performed on each patch to yield the corresponding $p\times p$ frequency coefficient array. The tensor is then reshaped such that the same coefficient across all patches forms each channel, and the channels are reordered by increasing coefficient (i.e., decreasing energy). After the reordering, the channels are truncated by $n$\%, so the lowest $n$\% of frequency coefficients (largest energies) are kept. This outputs highly compressed representations for the ICs and BCs, which are used as inputs for an RNN propagation branch that occurs completely in the frequency domain.

\textbf{Spatial branch.} 
The {spatial branch} follows a traditional ResNet \cite{he2016deep} architecture, in which the ICs and BCs each pass through separate convolutional encoders consisting of sets of convolutional blocks with residual connections. 
The inputs are downsampled with strided convolutions before entering the RNN propagation stage in the spatial domain.

\textbf{RNN propagation. } 
After compression, the representations enter the RNN propagation stage (\autoref{fig:architecture} (b)), in which the BCs are split into an additive ($B^{\rm bc}$) and multiplicative ($W^{\rm bc}$) component and combined with an IC-informed state matrix ($H_{t}$). 
The final output at each timestep is computed as $S_{t} = W^{\rm bc}H_{t}+B^{\rm bc}$. This method offers flexibility and efficiency in learning the dynamics of compressed simulations \cite{hennigh2017lat}. 
To predict the simulation state at each successive timestep, the previous hidden state $H_{t-1}$ is passed through a convolutional GRU (ConvGRU) along with the previous output $S_{t-1}$; for timestep 0, the initial state $H_0$ set to zero and ICs are used as inputs. This occurs in a recurrent manner until the final time $T$. Thus, for each timestep, the RNN propagation stage outputs $S_t$ which is then sent to a decoding step corresponding to the original frequency or spatial encoding. 
\begin{equation}
\begin{aligned}
    S_{0} &= u_{0}, \; H_0 = \mathbf{0}. \\
    H_t &= \text{ConvGRU}(S_{t-1}, H_{t-1}), \\
    S_t &= W^{\rm bc}H_{t}+B^{\rm bc},  \; t \in \{1, \dots, T\}.
    \label{eq:gru}
\end{aligned}
\end{equation}

The RNN propagation stage is duplicated across both frequency and spatial branches. 

\textbf{Latent grid generation.} 
After RNN propagation, the outputs are combined to form the latent grid. In the frequency branch, the output state  at each timestep from the RNN is converted back into the spatial domain. This is done by reshaping the frequencies from coefficients to patches, performing IDCTs, and then merging the patches to reconstruct the spatial domain. The output of the frequency branch is denoted as $O^{f}_{t}$. The representation in the spatial domain $O^{s}_{t}$ is then added with learnable weights $W^{o}_{t}$. Thus, the final output is computed as: $O_{t} = W^{o}_{t}O^{s}_{t} + O^{f}_{t}$.
These combined outputs $O_{t}$ for each timestep are used to form the spatio-temporal latent context grids. 
Finally, the multiple resolutions of grids are generated by upsampling the outputs $O_{t}$ using transpose convolutional blocks.


\subsection{Decoding step}
The multi-resolution latent context grids generated from the previous step are then used to query points input to the MLP. This decoding step follows the same principles as \cite{esmaeilzadeh2020meshfreeflownet}. 
Given a random query point $\mathbf{x}$ := $(x, y, t)$, $k$ neighboring vertices of the query point at each dimension are selected. Using these neighboring vertices, the final values of the context vector are then interpolated using Gaussian interpolation.
This process is repeated for each of the multi-resolution grids allowing the PINN framework to learn spatio-temporal quantities at multiple resolutions.

\subsection{Physics-informed loss}
The MLP outputs predictions that are then subject to the loss function determined by the ICs, BCs, and the PDEs. The losses are backpropagated through the entire combined decoding and latent grid network and minimized via stochastic gradient descent. This end-to-end training allows our two-branch convGRU model to learn accurate time-evolution of the spatial and frequency domains in complex physical problems.

\section{Experiments}
\label{sec:experiments}

We compare our model (\textbf{RNN-SpDCT}) against several other neural solver architectures using the 2D Taylor-Green vortex problems. 
This problem is commonly used to test and validate spatial and temporal accuracy of both traditional and ML-based fluid solvers. 
We compare against PINN-based models and use the ICs, BCs, and PDE constraints for all comparing models. We used a single Tesla V100 16G or 32G for all experiments. The patch-wise DCT use more GPU memory than other models.

\textbf{Baseline models. } 
We compare our proposed model against several PINN-based approaches: MLP-PINN, RNN-S, RNN-pDCT, and RNN-SfDCT. 
All comparing models contain the RNN-propagation and decoding steps except for MLP-PINN and all use physics informed loss explained in \autoref{sec:methods}. All use the same hyper-parameters as our model except for learning rate and decay steps.
\textbf{MLP-PINN}: a traditional MLP-based PINN solver used as a default model from \verb+SimNet+ \cite{hennigh2021nvidia}. 
\textbf{RNN-S}: a PINN solver with a latent grid network consisting of a single spatial branch (ResNet). 
\textbf{RNN-pDCT}: a PINN solver with a latent grid network consisting of a single frequency branch (DCT).
\textbf{RNN-SfDCT}:a PINN solver with a latent grid network consisting of both spatial and frequency branches. The frequency branch in this model applies DCT/IDCT to the full input, foregoing the patching, coefficient channel reordering, and truncation steps. 

\subsection{Implementation Details}
\textbf{Experimental setup. } 
The input size is set to $64 \times 64$. We used $2\pi$ seconds for both training and testing. The number of interpolation points $k$ used in the decoding step is $3$, and the truncation ratio $n$ is fixed to $25\%$. All models are trained with an Adam optimizer~\cite{kingma2014adam}. The starting learning rate is searched between \num{1e-4} and \num{4e-4}. We found that patch-wise models need lower learning rate \num{1e-4} and the others with higher learning rate \num{4e-4} with $0.95$ decay rate and different decay steps: 8000 steps for patch-based models, 2000 steps for other models.

\textbf{Network architecture. } 
The patch size $p$ in the patch-based DCT models is set to 8. The number of (encoding) residual blocks are 2 for the spatial branch and 1 for the frequency branch. In the spatial branch, 4 additional residual blocks are used with stride 2 for downsampling . There is no downsampling layer for the frequency branch. All convolutional layers have a filter size of $3 \times 3$, and there are two RNN propagation layers. The number of upsampling layers is searched between 1 and 4, and the reported numbers are with 4 layers. 


\subsection{Taylor-Green vortex}
The Taylor-Green vortex describes a decaying vortex flow which follows a special case of the Navier-Stokes equations \cite{taylor1937mechanism}. The incompressible Navier-Stokes equations in 2D are
\begin{equation}
  \begin{aligned}
    \partial_{x}u + \partial_{y}v & = 0 \\
    \partial_{t}u + u \partial_{x}u + v \partial_{y}u & = -\partial_{x}\rho/\rho + \nu (\partial_{xx}u + \partial_{yy}u) \\
    \partial_{t}v + u \partial_{x}v + v \partial_{y}v & = -\partial_{y}\rho/\rho + \nu (\partial_{xx}v + \partial_{yy}v). \\
  \end{aligned}
\end{equation}
where $u$ and $v$ are the $x$- and $y$-velocities, respectively, $\nu\in \mathbb{R}_{+}$ is the kinematic viscosity, and $\rho$ is the density. 

The exact closed form solution for the Taylor-Green vortex over the domain $x \times y \times t\in [0,2\pi]\times[0,2\pi] \times[0,T] $ is
\begin{equation}
  \begin{aligned}
    u & = \cos x \sin y F(t) \\
    v & = -\sin x \cos y F(t) \\
    p & = \frac{-\rho}{4}(\cos 2x + \cos 2y) F^2(t) \\
  \end{aligned}
\end{equation}
where $F(t)=e^{-2 \nu t}$ and $p$ is the pressure.


\subsection{Results}

\begin{figure}
  \centering
  \includegraphics[trim={4cm 1.3cm 2.5cm 0.3cm},clip, width=\linewidth]{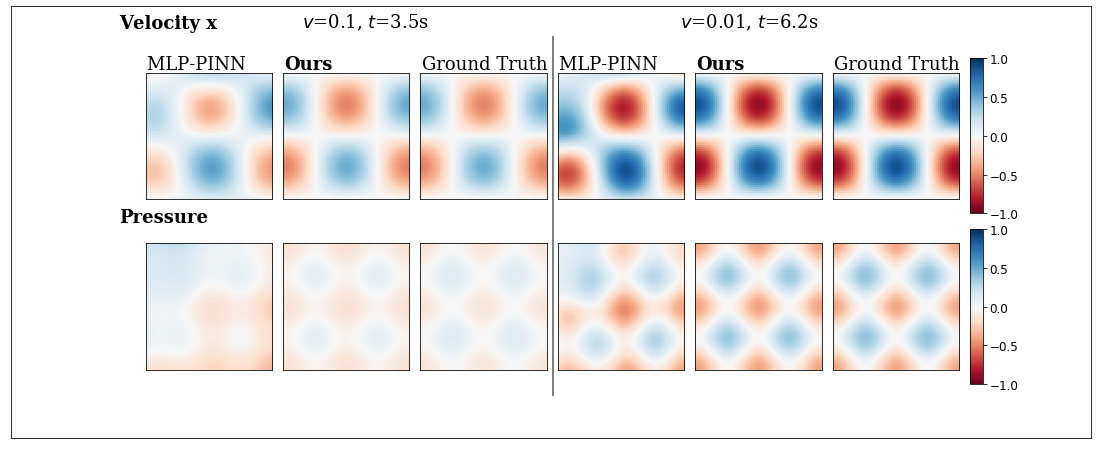}
  \caption{
    Visualization of the predictions on Taylor-Green vortex with the viscosity $\nu=0.1$ at around 3.5 seconds (left) and $\nu=0.01$ at around 6 seconds (right).}
  \label{fig:taylor_green}
\end{figure}

Table~\ref{tab:results} summarizes the performance of our model compared to the other PINN baselines. RNN-SpDCT achieves the best performance for all values of vorticity. All RNN models achieve extremely accurate velocities compared to MLP-PINN. 
\autoref{fig:taylor_green} visualizes the predictions and compares with the analytical solution. The model produces much more accurate predictions for longer time steps (up to $2\pi$ seconds) compared to MLP-based PINNs.

\begin{table}
  \caption{Quantitative comparisons on Taylor-Green vortex. The model is trained and tested for $2\pi$ seconds. The averaged mean squared error (MSE) over 10 uniformly sampled time steps is reported. $\nu$ is the kinematic viscosity of the fluid. F and S indicate frequency and spatial branches. FullDCT applies DCT to the entire input. 
  The numbers are in the magnitude of $10^{-2}$. 
  }
  \label{tab:results}
  \renewcommand{\tabcolsep}{1pt}
  \centering
  \footnotesize
  \begin{tabular}{l|c|c|cc|cc|cc}
    \toprule
    \multirow{3}{*}{Model Name}       &  \multirow{3}{*}{Branch} & \multirow{3}{*}{DCT type} & \multicolumn{6}{c}{Taylor-Green Vortex} \\ 
                                   & & & \multicolumn{2}{c|}{$\nu=1.0$} & \multicolumn{2}{c|}{$\nu=0.1$} & \multicolumn{2}{c}{$\nu=0.01$}  \\

                       & & &  velocity & pressure  &  velocity &  pressure &  velocity & pressure \\ 
                        
    \midrule
    MLP-PINN   & - & - & 0.033 & 5.910 &  1.769 &    0.782 & 0.824 & 0.522 \\
    \midrule
    RNN-S  & S &  -  & 6.683e-8 & 0.075 & 2.975e-7 & 0.138  & 2.527e-7   & 0.020 \\ 
    RNN-pDCT & F & patch & 1.979e-6 & 0.172 & 5.957e-7 & 1.383 & 8.804e-7  & 0.508 \\ 
    RNN-SfDCT & S+F & full  & 9.171e-8 & 1.177 & 2.961e-7 & 0.301 & 7.015e-6   & 0.018 \\ 
    \midrule
    \textbf{RNN-SpDCT} & S+F & patch & 1.408e-7 & \textbf{0.044} & 3.107e-7 & \textbf{0.101} & 1.328e-6 & \textbf{0.012} \\
    \bottomrule
  \end{tabular}
\end{table}

\begin{figure}
  \centering
  \includegraphics[trim={0.8cm 0.5cm 0.cm 0.5cm},clip,width=0.48\linewidth]{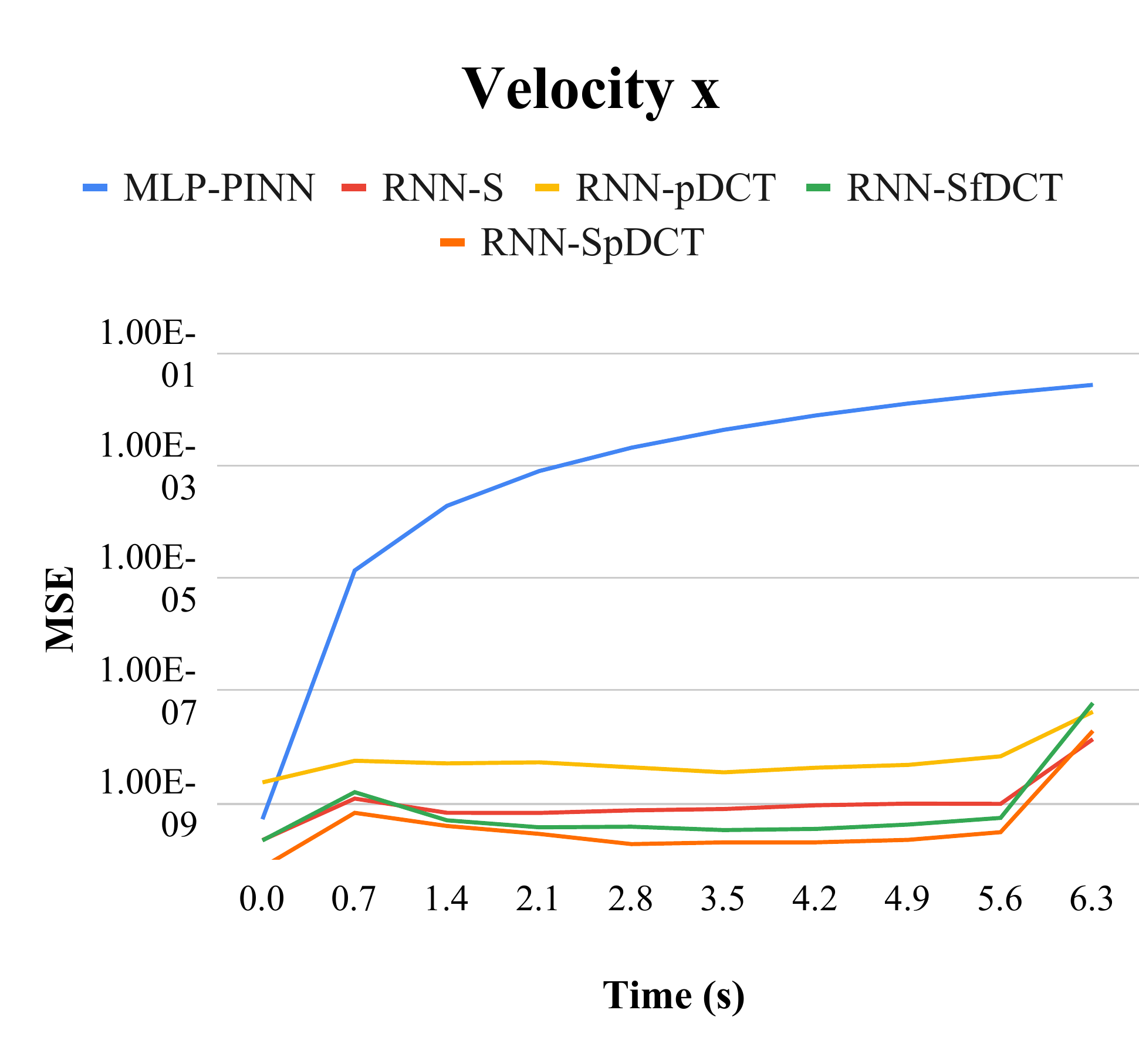}
  \includegraphics[trim={0.8cm 0.5cm 0.cm 0.5cm},clip,width=0.48\linewidth]{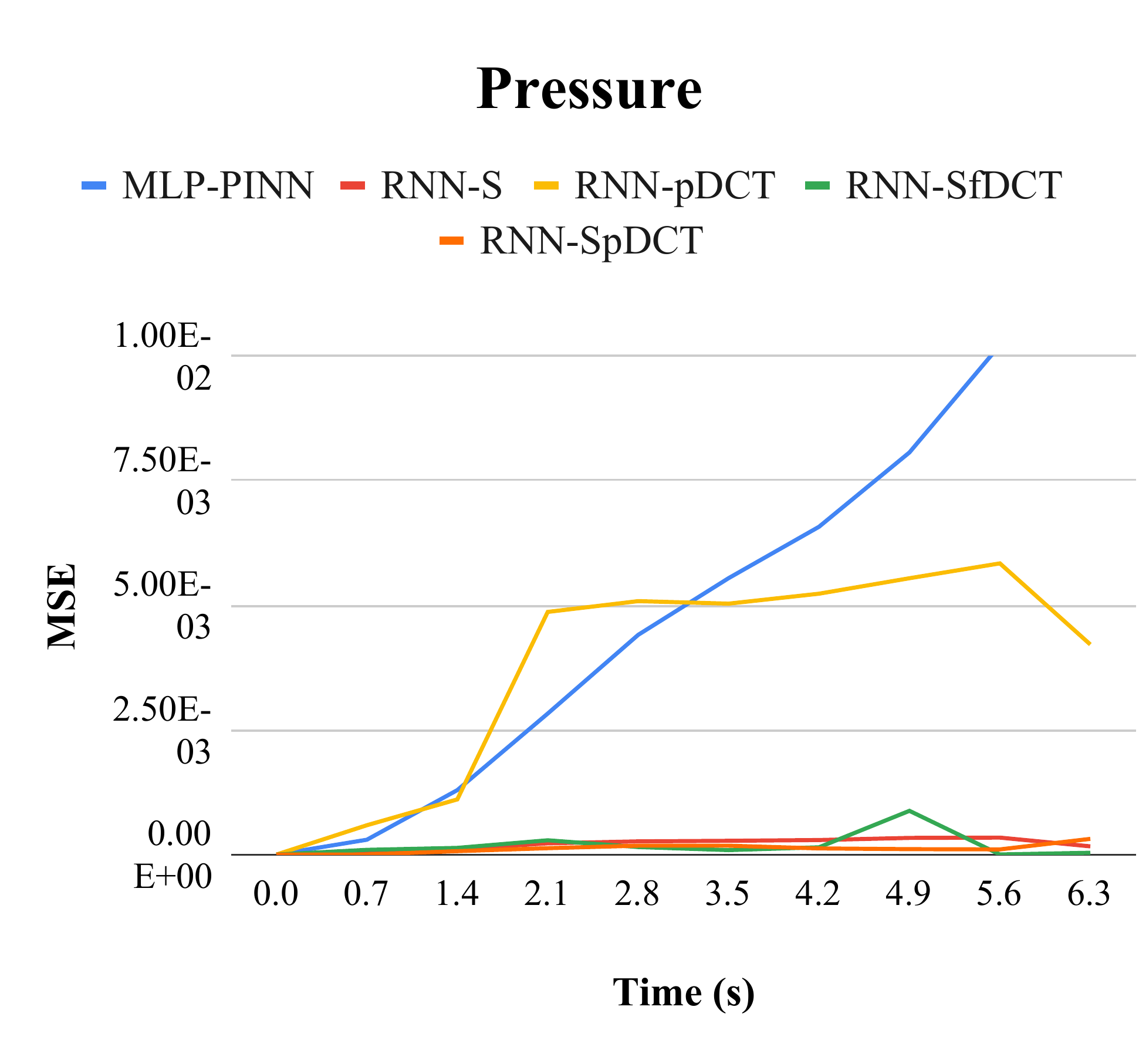}
  \caption{MSE comparisons over time. The MSE of the five models with respect to ground truth are shown over time for the Taylor-Green vortex problem. The x-velocity (left) is approximated with much lower error throughout long time evolution for the RNN models compared to the baseline MLP-PINN. The pressure (right) is also represented with very low error with several RNN models. In both cases, RNN-SpDCT achieves the best overall performance. 
  }
  \label{fig:taylor_green_time}
\end{figure}


\section{Conclusion}
We presented a novel extension to the PINN framework designed especially for time-dependent PDEs. Our model utilizes RNNs and DCTs to generate a multi-resolution latent context grid to condition the traditional MLP PINN architecture. We demonstrated that our model can accurately predict the solution functions in Taylor-Green vortex simulations (especially for pressures) and achieve state-of-the-art results. 
Future directions include experiments on more complex problems, higher dimensions, and longer time evolution. 


%
%
%
%
\bibliographystyle{splncs04nat}
\bibliography{biblio}
\newpage

\end{document}